\documentclass[12pt,preprint]{aastex}
\usepackage{graphicx}
\usepackage[dvips]{color}
\usepackage{txfonts}
\usepackage{amssymb}
\usepackage{natbib}
\usepackage{epstopdf}

\begin{document}
\title{Observational signatures of convectively driven waves in massive
  stars}
\author{C. Aerts}
\affil{Instituut voor Sterrenkunde, KU\,Leuven, Celestijnenlaan 200D, 3001
  Leuven, Belgium}
\affil{Department of Astrophysics/IMAPP, Radboud University Nijmegen,
             6500 GL Nijmegen, The Netherlands}
\author{T.M. Rogers}
\affil{Department of Mathematics and Statistics, Newcastle University,
  UK}
\affil{Planetary Science Institute, Tucson, AZ 85721, USA}





\begin{abstract}
  We demonstrate observational evidence for the occurrence of
  convectively driven internal gravity waves (IGW) in young massive O-type stars observed with high-precision CoRoT space
  photometry. This evidence results from a comparison between velocity spectra based on 
2D hydrodynamical simulations of IGW in a differentially-rotating massive star and
the observed spectra.We also show that the velocity spectra 
  caused by IGW may lead to detectable line-profile variability 
  and explain the occurrence of macroturbulence in the observed line profiles of OB stars.  Our findings
  provide predictions that can readily be tested by including a sample of bright
  slowly and rapidly rotating OB-type stars in the scientific programme of the
  K2 mission accompanied by high-precision spectroscopy and their
  confrontation with multi-dimensional hydrodynamic simulations of IGW for various masses and ages.
\end{abstract}


\section{Introduction}

The existence of high-order gravity-mode (g-mode) oscillations in Slowly Pulsating B
(SPB) stars, which are core-hydrogen burning stars with mass between roughly 3
and 7\,M$_\odot$ \citep[SPBs hereafter,][]{Waelkens1991} was established more
than two decades ago, prior to understanding their excitation
mechanism. We now know that these oscillations are driven by the
heat-mechanism associated with the opacity
bump due to iron-group elements in the stellar envelope
\citep{Dziembowski1993,Gautschy1993}. However, the detection of g-mode
period spacings associated with those standing waves as predicted from theory
\citep{Tassoul1980} and required for asteroseismology of such stars,
remained impossible from the ground, even after carefully
planned long-term dedicated campaigns
\citep{Aerts1999,DeCatAerts2002}.  This is partly
 due to the low amplitudes of g-modes and
is also made difficult because the
periods of gravity modes in massive stars are of order days and these
timescales have
strong contamination by daily aliases in the amplitude spectra of ground-based
photometry and high-resolution spectroscopy.  Gravity-mode asteroseismology of SPBs only saw its birth due to the
uninterrupted high-precision space photometry assembled with the CoRoT and {\it
  Kepler\/} missions. These data led to the detection of period spacings caused
by dipole modes of consecutive radial order and offered the mode identification
necessary for seismic modelling
\citep{Degroote2010,Papics2012,Papics2014,Papics2015}.

While heat-driven g-mode oscillations in SPBs are strictly periodic and
have a well-known and quantifiable effect on observed time-series photometry and
spectral line-profile variations \citep[e.g.,][]{DeCatAerts2002,Aerts2014}, the
effect of waves excited stochastically by core convection
\citep[][]{Samadi2010,Belkacem2010,Shiode2013} on such
observations is relatively unknown.  This lack of observational
diagnostics connected with IGW is particularly relevant in the context
of the variability of O-type stars, in which heat-driven modes are not
excited, while their large convective cores likely drive IGW
efficiently.  This work is therefore focused on the search for IGW driven by core convection in photometric and
  spectroscopic observations of a few carefully selected observed 
O-type stars.  Such signatures are important
  because the existence of these waves could point to 
  enhanced angular momentum transport and chemical mixing and hence,
  guide inclusion of these processes into future theoretical models.  
\begin{figure*}
\begin{center}
\rotatebox{270}{\resizebox{5.8cm}{!}{\includegraphics{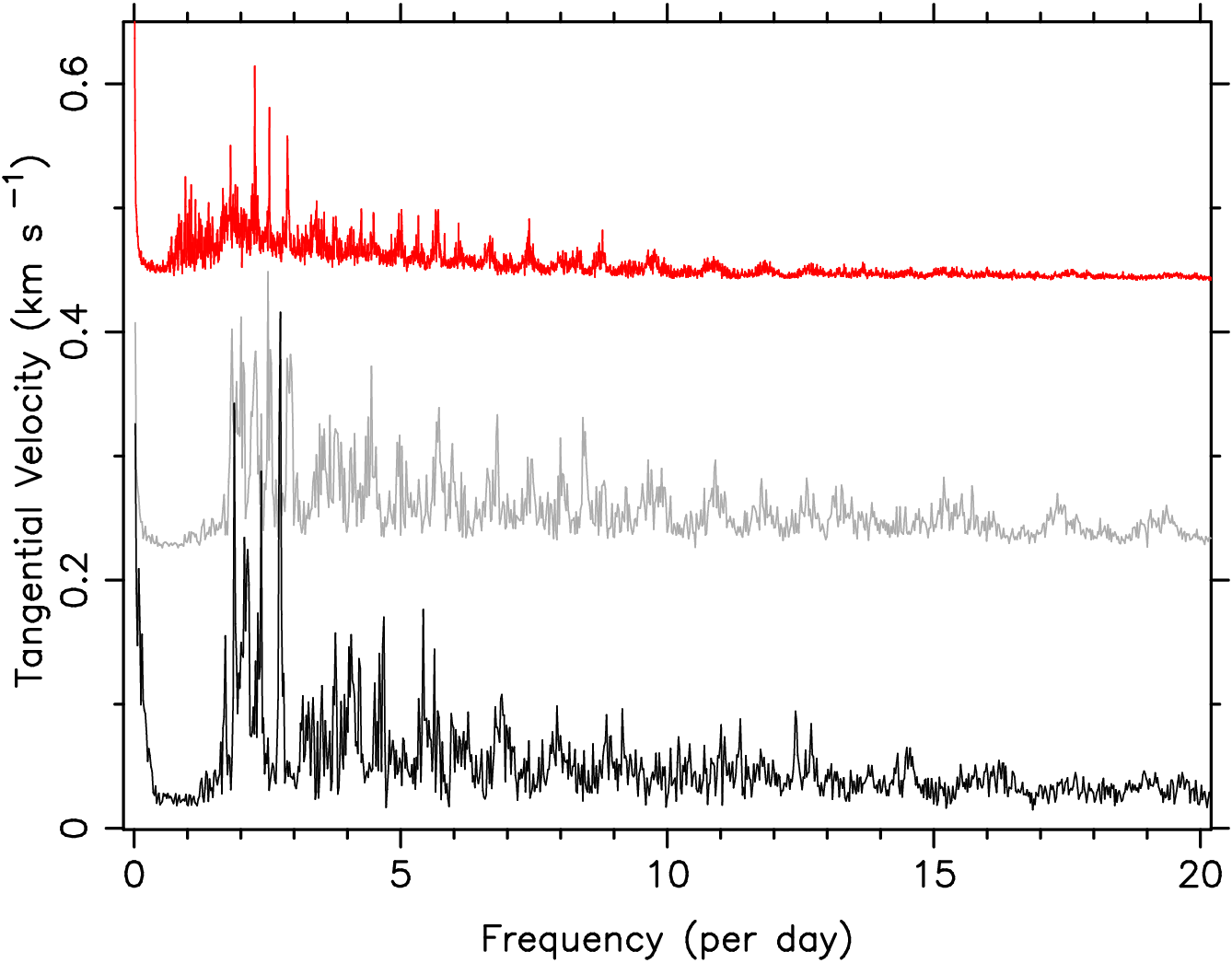}}}\hspace{0.5cm}
\rotatebox{270}{\resizebox{5.8cm}{!}{\includegraphics{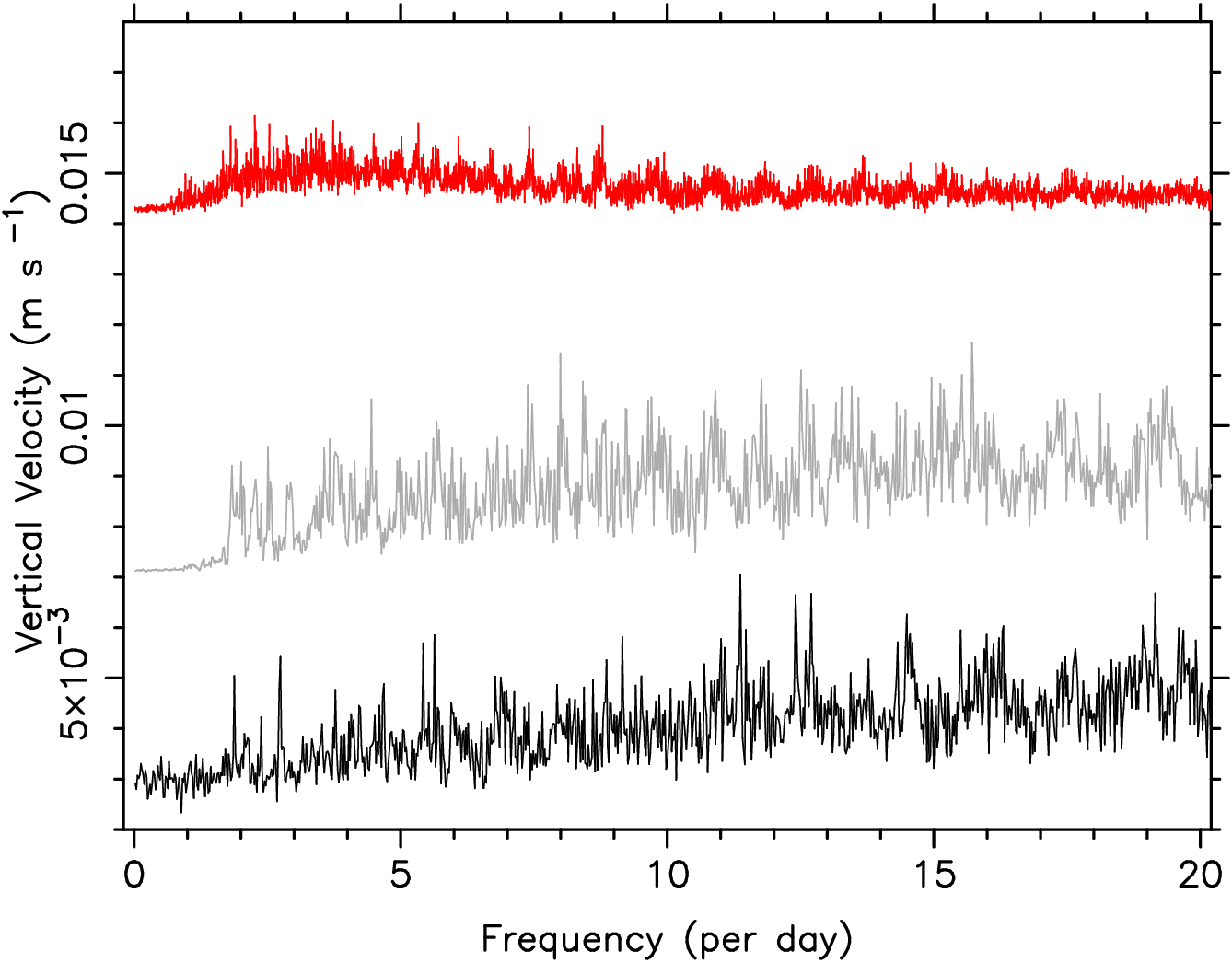}}}
\end{center}
\caption{Velocity spectra due to IGW for three cases: a non-rotating star
  (black), a rigidly rotating star with initial rotational frequency of 1.1\,d$^{-1}$
  (grey), and a differentially rotating star with the core rotating 1.5 times
  faster than the envelope for an initial surface rotational frequency of 0.275\,d$^{-1}$
  (red), shifted along the $y$-axis for visibility purposes. 
Left: tangential velocity; right: vertical velocity.}
\label{Tami}
\end{figure*}

\section{Modeling convectively driven waves}

There have been relatively few theoretical predictions for the spectra
and amplitudes of IGW excited by core convection in massive stars.
On the one hand, \citet{Browning2004} provided the first 3D simulations
of the core convection  for a 2\,M$_\odot$ star but these only
covered the inner 30\% in radius and hence, omitted much of the wave
propagation region. More recent
theoretical models do consider the entire radius of the star but are limited
to one-dimension (1D), neglect rotation and must make assumptions
about the nature of convection and convective-overshoot
\citep{Samadi2010,Shiode2013}.  
Generally those theoretical models assume 
Reynolds stresses in the convection zone that generate waves with a
predominant frequency given by the convective turnover frequency and
with  frequency and wavelength spectra dictated by the assumed properties of the
turbulence.  The amplitudes of the waves are determined by the
efficiency with which energy is transferred from convection to waves,
and is highly dependent on assumptions about the convective-radiative
interface.  Despite these shortcomings, the theoretical spectra of IGW predicted by
\citet{Samadi2010} and \citet{Shiode2013} are consistent with the
observed frequency ranges of variable OB stars.  However, their
amplitudes vary significantly and appear to be inconsistent with observations.
For example, the amplitudes predicted in \citet{Shiode2013} are more
than an order of magnitude lower than those predicted in
\citet{Samadi2010} and even those are too small to explain the detection of stochastically excited
gravito-inertial modes in HD\,51452 by \citet{Neiner2012}.  These
theoretical models also
assume that the star is non-rotating and it was shown by
\citet{Mathis2014} that the Coriolis force has a severe effect on the
stochastic excitation of waves and hence, their
surface amplitudes.  In reality, the waves in stars are mixed
gravito-inertial waves and this should be accounted for in
theoretical predictions of surface wave amplitudes\footnote{For ease
  of presentation in the following comparisons we will use IGW
  generically to mean convectively driven mixed gravito-intertial
  waves, sometimes referred to as GIW.}.  Beyond their effect on photometric and
spectroscopic observations, these mixed type waves
likely lead to substantial angular momentum transport
\citep{Rogers2013} and may be responsible for the Be star phenomenon
\citep{Ando1986,Rogers2013,Lee2014},  which could not be fully captured in
earlier 1D analytic models.  Clearly, observational
constraints on the detection of IGW, as well as more realistic multi-D
predictions of their observational signature, would be highly beneficial.

Here, in order to make comparisons between theoretical predictions of IGW in massive
stars and observations, we use the velocity spectra from the
two-dimensional (2D) numerical
simulations of convectively driven waves in a 3\,M$_\odot$ star,
described in \citet{Rogers2013}. While having many shortcomings
(dimensionality, degree of turbulence), these simulations provide a
more realistic benchmark for comparison because they self-consistently
generate the waves by convection, include all non-linearity and rotation. In order to carry out such
simulations some sacrifices had to be made.  Perhaps the most
relevant for the current work, are that the simulations used an enhanced thermal diffusivity for
numerical stability and considered only one mass.  In order to compensate for
the enhanced thermal damping of the waves on their journey to the
stellar surface, the waves were driven harder by using an
(equivalently) enhanced heat flux.  While this is not a perfect
representation of the waves, it is likely the best that can be done with
current computational resources.  Because of this shortcoming, wave
amplitudes produced by these simulations should be treated with some
caution.  Furthermore, the simulations only  
considered one mass and it is well known that the IGW frequencies
depend weakly on the mass of the star
\citep[e.g.,][]{Aerts2010,Shiode2013}.  Repeating the simulations for various masses and evolutionary stages is beyond
the scope of the current work, given that each 2D simulation takes
approximately a hundred
thousand processor hours for \textit{one} set of parameters
(mass, age, rotation, diffusion coefficients, luminosity). Such a simulation study will be the
subject of future research.  In order to compensate for the enhanced
thermal diffusion, the convective heat flux adopted in 
\citet{Rogers2013} was typically 10$^4$ times higher than
that of a 3\,M$_\odot$ star, so the predicted spectra of IGW might
appropriately resemble those of $\sim$\,30\,M$_\odot$ stars since the
wave spectrum is predominantly determined by properties of the convection, but this is yet to be
investigated.  In order to account for variations in mass
with respect to the 3\,M$_\odot$ and for the too high convective flux adopted
for the currently available numerical
simulations, we
apply one multiplicative factor (an unknown parameter) to scale the amplitudes
of the waves and another one to rescale frequencies for comparisons with real
stars, as discussed in more detail below. Despite these caveats, we expect that
the
overall characteristics and {\it shape\/} of the predicted frequency spectra
should offer a realistic representation of the effect of IGW on observables.

In Fig.\,\ref{Tami}, we show velocity spectra for three cases from \citet{Rogers2013}: a non-rotating
star, a rigidly rotating star with an initial rotational frequency of
1.100\,d$^{-1}$, and a differentially rotating star whose core rotates initially
with a factor 1.5 times faster than its envelope, the surface rotational
frequency being 0.275\,d$^{-1}$. These cases correspond with the simulations
labelled as U1, U8, and D11 in Table\,1 of \citet{Rogers2013}, to which we refer
for details. Using a radius from the prototypical SPB KIC\,10526294
with $M=3.25\,$M$_\odot$ and $R=2.215\,$R$_\odot$ \citep{Moravveji2015}, the two
rotating models give initial equatorial rotation velocities of 31 (non-rigid,
D11) and 123\,km\,s$^{-1}$ (rigid, U8).  The velocity spectra were
obtained after more than 100 rotational cycles (for U8 and D11), in a layer near the stellar
surface at 99\% in radius, and the tangential
velocity is an average over $\varphi\in [0,2\pi]$.  In
Fig.\,\ref{Tami}, it can be seen that the
vertical velocity is many orders of magnitude smaller than the tangential
velocity, as is expected for gravity waves. Fig.\,\ref{Tami} also clearly
illustrates that non-rigid rotation (D11) inside the star shifts the power to
lower frequencies compared to the rigidly or non-rotating
cases. Given that the few estimates of core-to-envelope rotation of massive stars
  point to nonrigid rotation with a factor 2 to 5 faster core than envelope
\citep[e.g.,][]{Aerts2003,Pamyatnykh2004}, 
we use D11 for comparisons with observations.

Beyond the uncertainties in wave amplitudes and frequencies discussed
above, it is not entirely clear how to convert
velocity spectra, of the type seen in Fig.\,\ref{Tami}, to brightness
variations, as observed.  To address this, we can use the {\it known amplitude ratio\/} of
photometric brightness variability due to heat-driven g-mode
oscillations in SPBs to velocity variations measured using spectral
line diagnostics in the same stars.  The case of convectively driven waves
is, of course, different as these waves have shorter lifetimes
and some are breaking leading to turbulence.  
On the other hand, the heat-driven prograde or retrograde
g modes of SPBs correspond with trapped internal
  resonant gravity waves.  The 2D simulations produce convectively
  driven resonant gravity waves with amplitudes in the same range as
  the non-resonant waves.  Therefore, in absence of any better
  conversion, we
  assume that the brightness fluctuation signatures due to velocity
  variations from IGW behave similarly to those from 
  heat-driven g modes in terms of amplitude ratios of photometric and spectroscopic observables
measured at one instance in time. Under this assumption, we use the carefully
{\it measured}
amplitude ratios of $(\Delta B)/(\Delta\!<\!v\!>)$ of g modes to convert tangential velocity spectra to
brightness spectra.  The
measured amplitude ratios for the g-modes in the single SPB
pulsators considered in \citet{DeCatAerts2002} have values between roughly 1 and
10\,mmag/(km\,s$^{-1}$), as deduced from the light curves in the $B$ band and
from the centroid velocity $<\!v\!>$ derived from high-resolution high
signal-to-noise time-series spectroscopy \citep[see Chapter\,6 in][for a
definition]{Aerts2010}. Below, we use these observational results to
scale wave amplitudes in the simulations for comparison with observations.

In the following sections, we first provide observational evidence of the occurence of IGW in
$\mu$mag space photometry of three young unevolved O-type stars measured with
the CoRoT satellite. We then predict
line-profile variations due to simulated IGW in order to evaluate if IGW can provide an
alternative explanation of macroturbulent spectral line broadening, in addition to
coherent g-mode oscillations \citep[e.g.,][]{Aerts2009,SSD2010}.

\section{Photometric signatures: power excess in space photometry of hot massive
stars}

\begin{figure}
\begin{center}
\rotatebox{270}{\resizebox{4.5cm}{!}{\includegraphics{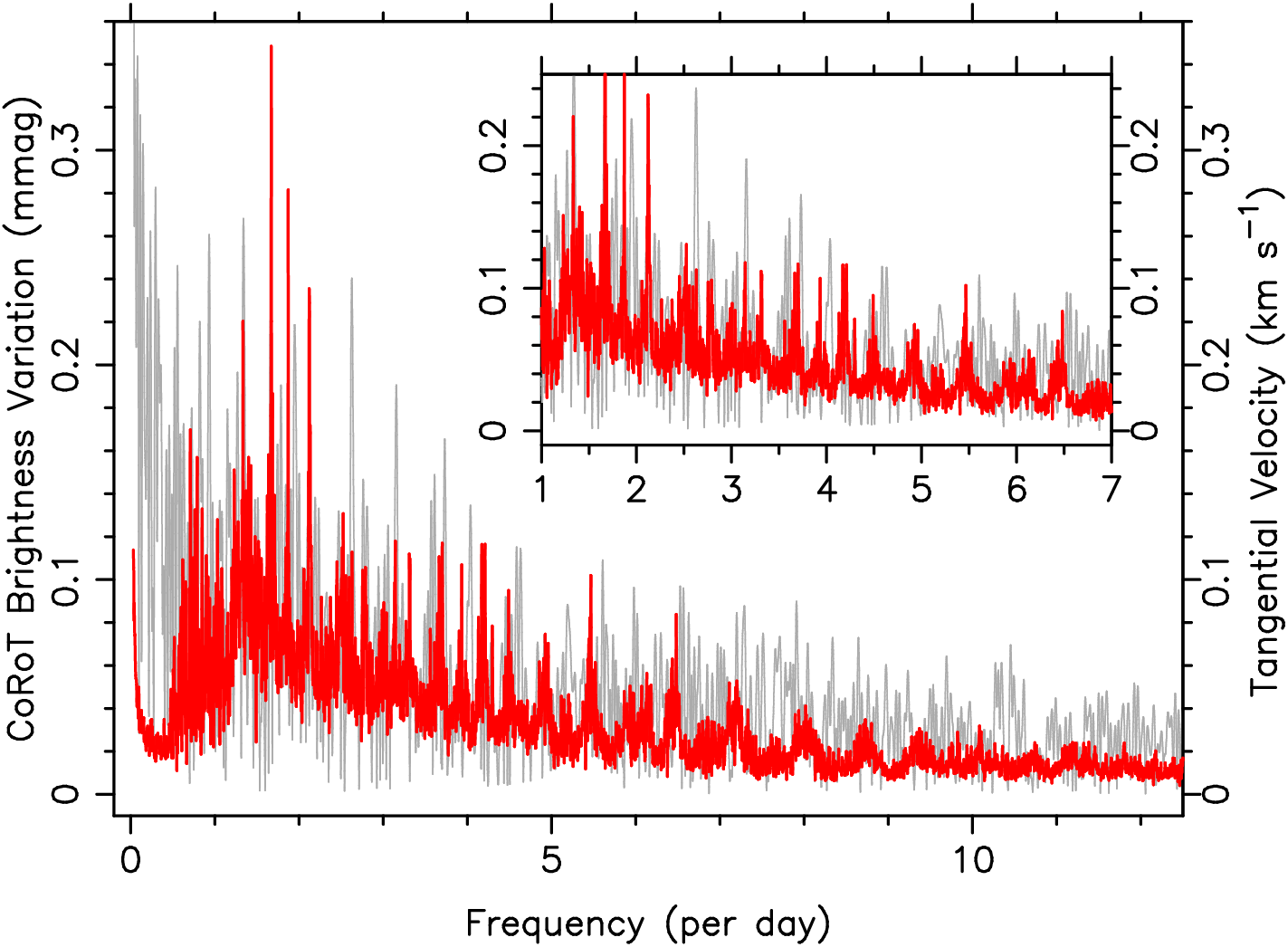}}}\\
\rotatebox{270}{\resizebox{4.5cm}{!}{\includegraphics{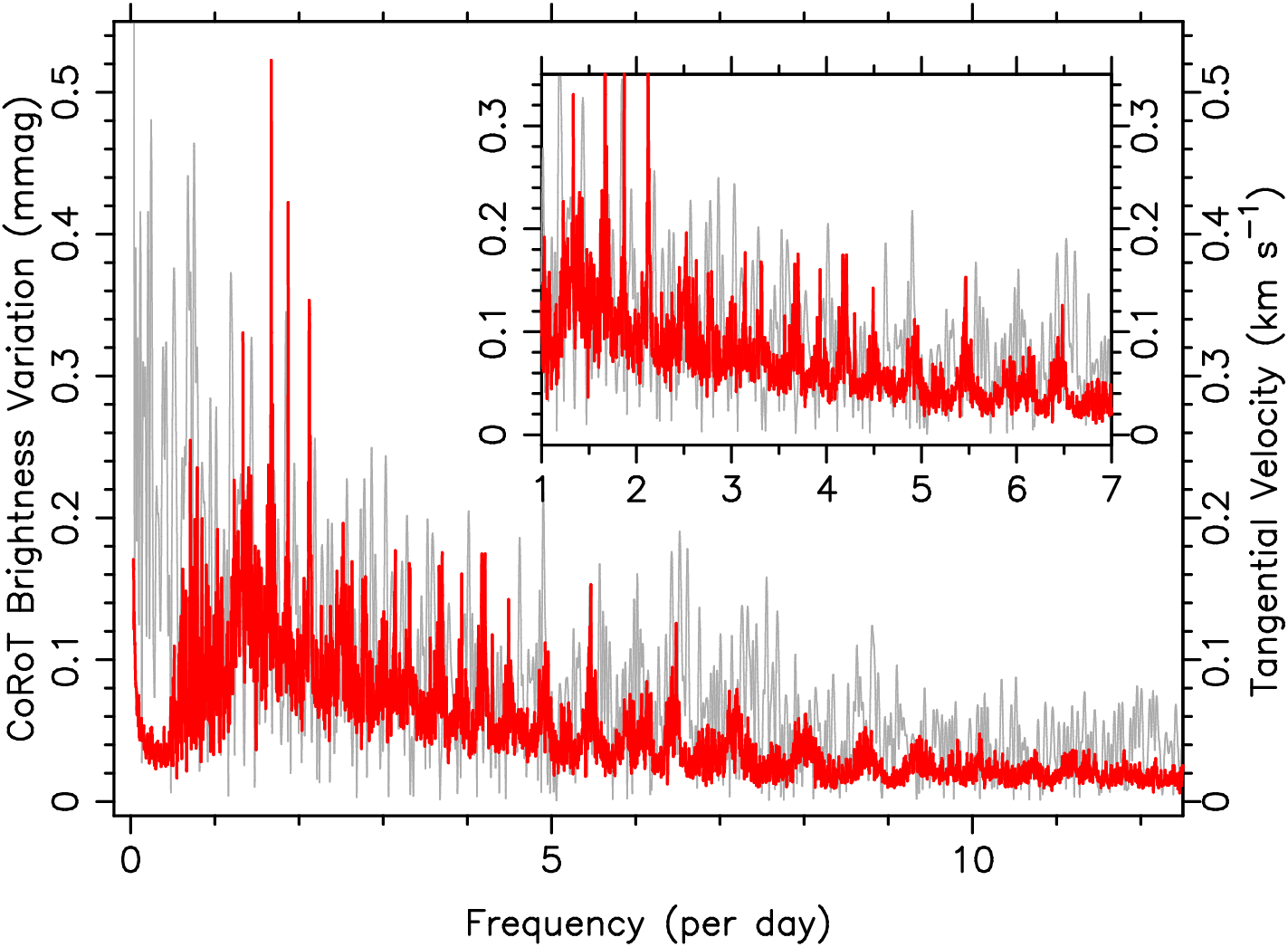}}}\\
\rotatebox{270}{\resizebox{4.5cm}{!}{\includegraphics{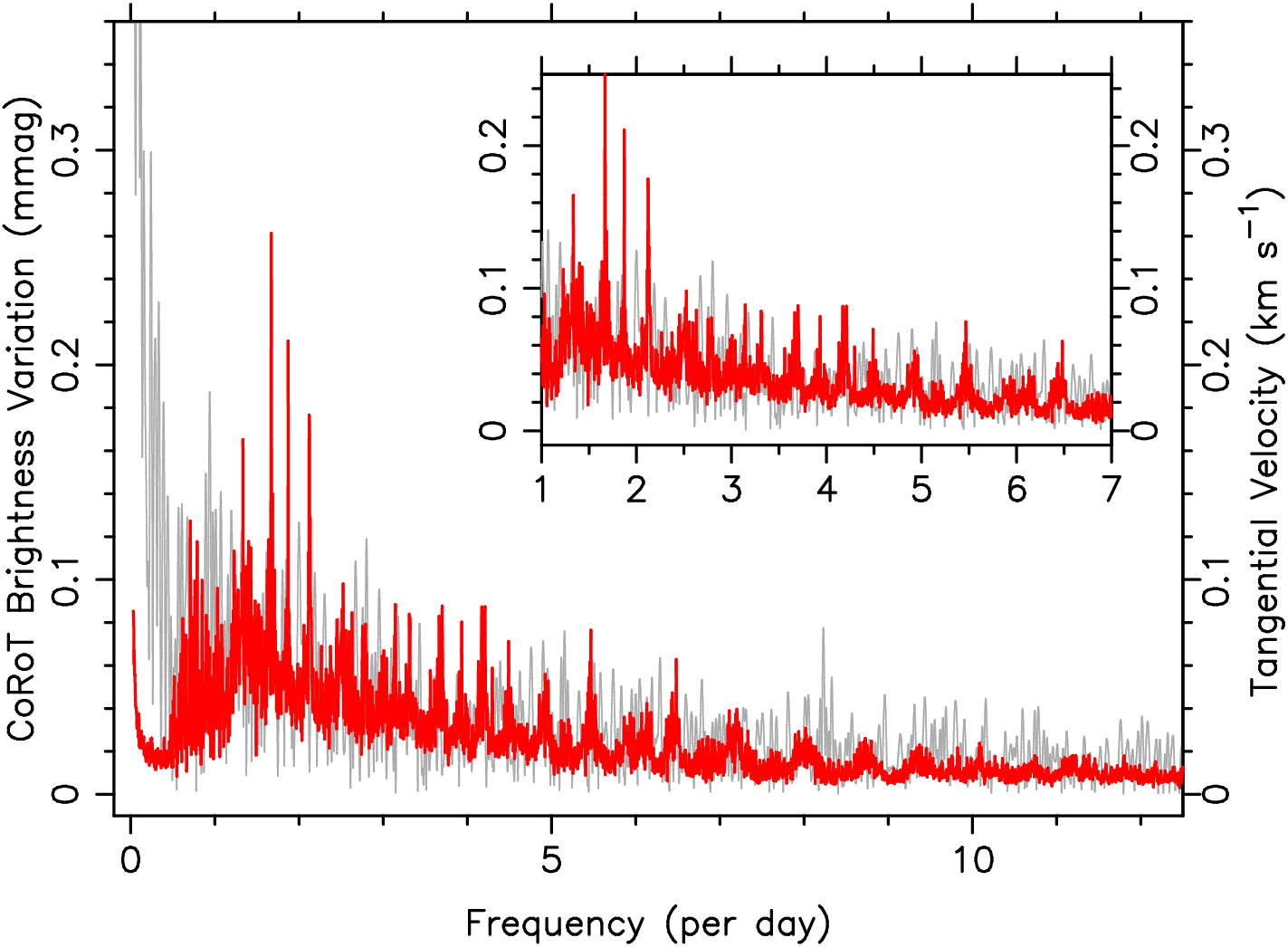}}}
\end{center}
\caption{Measured amplitude spectra (grey) overplotted with the predictions of
  IGW for case D11 (red) for HD\,46150 (top), HD\,46223 (middle), and HD\,46966
  (bottom). The ratios between brightness and tangential velocity variations
  used are 2, 3, and 1.5, from top to bottom.}
\label{Ostars}
\end{figure}

The MOST, CoRoT, and {\it Kepler\/} missions revealed a much larger diversity in
variability of OB-type stars at $\mu$mag level than anticipated prior to the era
of high-precision space photometry \citep[e.g.,][for an
update]{Neiner2011,Aerts2015}. While most of the variability is well
understood as due to heat-driven stellar oscillations, rotational modulation,
magnetic activity, mass loss, various sorts of binarity, or a combination of all
those processes, the ``red noise'' power excess found in the amplitude
spectra of three O stars, remains unexplained \citep{Blomme2011}.  

As is common in the study of heat-driven oscillations, we represent
the effect of IGW by means of amplitude spectra, where we focus on the
$\varphi$-averaged tangential velocity (see Figure 1).
Figure\,\ref{Ostars} shows the observed
amplitude spectra resulting from brightness measurements assembled by the CoRoT
mission for three O stars, with surface rotation frequencies of
0.144\,d$^{-1}$ for HD\,46150 (top panel),  0.249\,d$^{-1}$ for HD\,46223 (middle
panel), and 0.084\,d$^{-1}$ for HD\,46966 (bottom panel). These spectra are presented in mmag and have been
reproduced from \citet{Blomme2011}, who found them to reveal frequencies
connected with phenomena of short lifetimes (order of hours to
days). We compare those observed spectra (grey line in Fig.\,\ref{Ostars})
with the simulated IGW spectra shown in Fig.\,\ref{Tami} for the
non-rigidly rotating case D11 (red line in Fig.\,\ref{Ostars}). 
 
We have applied a correction factor of $\sim$0.75
to the frequencies to account for the factor $\sim$10 higher mass
between the observed and simulated stars, following \citet[][their Table\,1]{Shiode2013}.  
We realise that the propagation cavity of the IGW changes somewhat as a function of mass
and age, which may imply that such a simple scaling is not optimal. 
On the other hand, the spectrum of IGW is dominantly determined by the
convective flux and less so by the shape of the Br\"unt-V\"ais\"al\"a
frequency. For this reason, we consider a simple scaling to be the best
approach while awaiting future 2D/3D simulations for various stellar masses and
evolutionary stages.

We further applied a factor
of 2, 3, and 1.5, respectively to the IGW amplitude spectra (shown in
red in Fig.\,\ref{Ostars}) to convert
between simulated tangential velocity and brightness, 
consistent with the observed range of amplitude ratios from
$(\Delta B)/(\Delta\!<\!v\!>)$ for the heat-driven modes of SPBs. The
lowest frequencies, below 1\,d$^{-1}$, are not reproduced by the
simulations.  This could be due to too low a stellar mass and/or too low
level of differential rotation and/or the too simplified scalings we have
  adopted.
However, the overall resemblance is good in terms of
frequency structure. This interpretation of the ``red
noise'' power excess as due to IGW is currently the only theoretical
explanation that provides amplitude
spectra whose {\it shape\/} is 
in agreement with those  observed in the three O stars.

\section{Spectroscopic signatures: macroturbulence in spectral line 
profiles of hot massive stars} 

\begin{figure*}
\begin{center}
\rotatebox{270}{\resizebox{5.8cm}{!}{\includegraphics{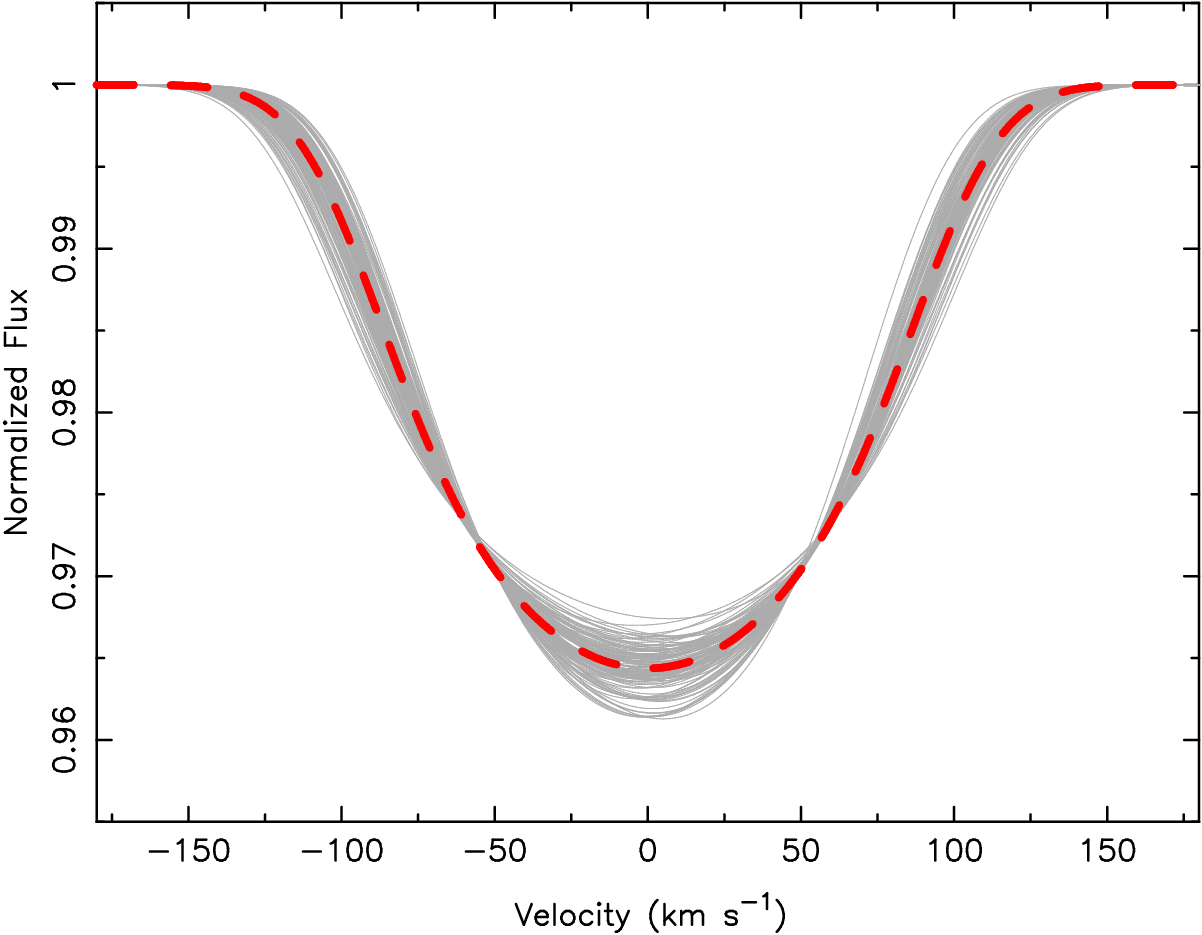}}}\hspace{0.5cm}
\rotatebox{270}{\resizebox{5.8cm}{!}{\includegraphics{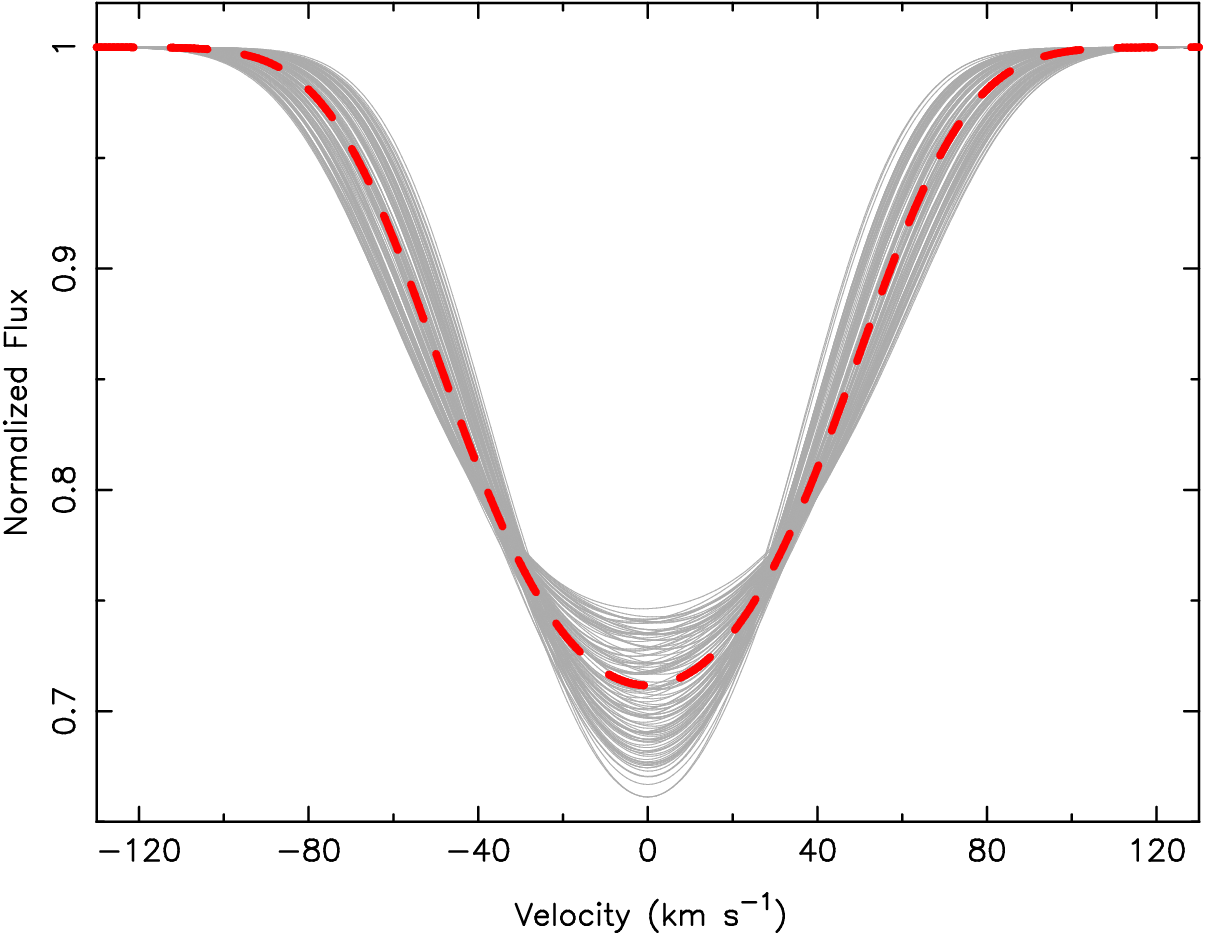}}}
\end{center}
\caption{Simulated line profile variations for the O\,III\,\,5922\AA\ line  of HD\,46150 assuming IGW (left
panel) and for the 
Mg\,II\,4481\AA\ line of the SPB KIC\,7760680, based on its
dipole heat-driven modes detected and identified  in the {\it Kepler\/} data
(right panel). The red dashed lines are for rotational broadening alone.}
\label{LPVs}
\end{figure*}
High resolution time series spectroscopy of confirmed g-mode
  pulsators show line-profile
  variations (LPVs) of a complex nature but such
  data is only available for a handful of SPBs \citep[e.g.,][]{DeCatAerts2002,Papics2012}.  
In absence of systematic time-resolved LPV
studies of gravity-mode pulsators among OB-type stars, \cite{Aerts2009} performed simulations to
show that the collective effect of numerous \textit{heat-driven\/} g-modes leads to
LPVs that mimic the effect of a macroturbulent velocity ($v_{\rm
  macro}$).   More recently, it was shown from observed LPVs that even one
mode or spot can give rise to spectral line broadening in B stars
consistent with considerable $v_{\rm macro}$ values
\citep{Aerts2014}. These findings indicate
that pulsational velocities due to heat-driven g-modes could explain the
occurrence of macroturbulence along most of the upper main sequence,
except for the highest-mass core-hydrogen burning O stars which do not show
heat-driven modes \citep{SSD2015}.  
However, 
they do show a power excess of frequencies with short
lifetime (Fig.\,\ref{Ostars}), which we demonstrated may be the signature of IGW.  It is then
reasonable to investigate whether IGW can lead to LPVs in these stars.

Using stellar parameters and frequencies
of HD\,46150 \citep{Blomme2011,Martins2015} and velocity fluctuations given in the D11
simulation (scaled by a factor of 2 as in the red curve in
the upper panel of Fig.\,\ref{Ostars}) we
produced  LPVs for its O\,III\,5922\AA\ line
 following the method outlined in \citet[][Chapter 6]{Aerts2010}.
  In doing so, we used the radial and tangential velocity fluctuations
  produced in the simulations (Fig.\,\ref{Tami}) as radial and
  tangential amplitude and we adopted a longitudinal and latitudinal dependence according to dipolar
  prograde spherical harmonics seen under an inclination angle of
  60$^\circ$.  This procedure produced the LPVs shown in the
    left panel of
  Fig.\,\ref{LPVs}.
Since there are no observed LPVs for these O-type stars,
for comparison we show simulated LPVs for the SPB star KIC\,7760680,
representative of its 
Mg\,II\,4481\AA\ line 
and based upon its identified
heat-driven modes detected in the {\it Kepler\/} data. In order to generate these
LPVs, we have taken the frequency,
amplitude, and phase values from Table\,1 in \citet{Papics2015}. The resulting
LPVs are shown in
the right panel of Fig.\,\ref{LPVs}. We find that the velocity fields due to
IGW that give good agreement with the CoRoT photometric data
(Fig.\,\ref{Ostars}) also lead to LPVs that are quite
similar to those produced by heat-driven gravity modes of KIC\,7760680 and that the
spectral line wings of both stars are predicted to be broadened compared to the
case of rotational line broadening alone, indicative of macroturbulence. 
Of course, we have presented just one set of simulated line
profiles based on particular assumptions about the nature of IGW in terms of
amplitude and mode type, but these simulations are in agreement with the need
for macroturbulence to fit the profiles of HD\,46150, as found by
\citet[][$v_{\rm macro}\simeq\,38\,$km\,s$^{-1}$]{Martins2015}.  Our simulations
also show that the anticipated LPVs resulting from the chosen angular
  velocity dependence and inclination angle are tiny in velocity space ($x$-axis
  in Fig.\,\ref{LPVs}) while better visible in relative flux changes ($y$-axis),
  i.e.\ the profiles remain fairly symmetric and fluctuate in relative flux
  level.  So actual future detection of time-variability in the
  O\,III\,\,5922\AA\ line of HD\,46150 will require very high-resolution, high
  signal-to-noise time series spectroscopy in order to observe the fluctuations
  in the broadening of the spectral line over time, with respect to a time-averaged
  profile.

\section{Discussion}

We have shown that the 2D simulations of IGW by \citet{Rogers2013} give rise to
averaged tangential velocity spectra in agreement with measurements of ampitude
spectra of flux variations in three mid-O type stars that do not exhibit
heat-driven oscillations. 
Our simulations of LPVs based on the velocity spectra due to IGW reveal that
such waves can give rise to detectable time-dependent 
line broadening similar to that 
observed in SPBs. IGW thus offer a viable explanation for the occurrence of
macroturbulence in the hottest main-sequence stars of spectral type O, 
which are not subject to heat-driven oscillation modes and/or strictly periodic
rotational modulation due to surface spots.  Given the shortcomings of
the numerical simulations, it is remarkable that they match the
observations within factors of two in both frequency and
amplitudes. This provides hope that numerical
simulations may be able to capture the gross properties of waves in stars
and that more sophisticated simulations will be able to make
comparisons more robust.

Given our results, it is worthwhile to search for signatures of IGW in the
residual light curves prewhitened for the detected
heat-driven mode frequencies for all OB stars observed by the {\it Kepler\/}
and CoRoT missions, keeping in mind the recent finding that many of the
frequencies found in the amplitude spectra of gravity-mode pulsators
are combination frequencies (possibly due to nonlinear mode coupling)
of heat-driven modes
\citep{Papics2015,Kurtz2015}.  However, it should be 
possible to distinguish between those combination frequencies
generated by heat-driven g-modes
and the signatures of IGW since IGW have short lifetimes leading to
frequency features with broad wings in the Fourier transform, while
combinations frequencies due to heat-driven modes with infinite lifetimes reveal delta-peaks in
the frequency spectra. While we provide this interpretation of the
nature of macroturbulence in O stars, an independent spectroscopic
study on the nature of macroturbulence has been initiated by \citet{SSD2014}.

In the near future, we also plan to gather new photometric data with the refurbished
{\it Kepler\/} mission, baptised K2 \citep{Howell2014}, accompanied by
high-resolution high-precision spectroscopic time series.  Such a
combination of data would allow us to test our findings and to provide a large enough sample of OB-type stars in
all evolutionary phases to test the occurrence of IGW across the evolution of
the most massive stars.

\bibliographystyle{apj}

\begin{thebibliography}{}


\bibitem[\protect\citeauthoryear{Aerts}{2015}]{Aerts2015} Aerts C., 2015, IAUS,
  307, 154

\bibitem[\protect\citeauthoryear{Aerts et 
al.}{2009}]{Aerts2009} Aerts C., Puls J., Godart M., Dupret M.-A., 2009, A\&A, 508, 409 

\bibitem[Aerts et al.(2010)]{Aerts2010} Aerts, C., Christensen-Dalsgaard, J., \&
  Kurtz, D.\ W.\ 2010, Asteroseismology, Springer, Heidelberg

\bibitem[\protect\citeauthoryear{Aerts et 
al.}{1999}]{Aerts1999} Aerts C., et al., 1999, A\&A, 343, 872 

\bibitem[Aerts et al.(2003)]{Aerts2003} Aerts, C., Thoul, A., 
Daszy{\'n}ska, J., et al.\ 2003, Science, 300, 1926 


\bibitem[\protect\citeauthoryear{Aerts et al.}{2014}]{Aerts2014} Aerts C.,
  Sim{\'o}n-D{\'{\i}}az S., Groot P.~J., Degroote P., 2014, A\&A, 569, A118

\bibitem[\protect\citeauthoryear{Ando}{1986}]{Ando1986} Ando,H., 1986,
  A\&A, 163,97

\bibitem[\protect\citeauthoryear{Belkacem, Dupret, 
\& Noels}{2010}]{Belkacem2010} Belkacem K., Dupret M.~A., Noels A., 2010, A\&A, 510, A6 

\bibitem[Blomme et al.(2011)]{Blomme2011} Blomme, R., Mahy, L., Catala, C., et
al.\ 2011, A\&A, 533, A4


\bibitem[Browning et al.(2004)]{Browning2004} Browning, M.~K., Brun, 
A.~S., \& Toomre, J.\ 2004, \apj, 601, 512 


\bibitem[\protect\citeauthoryear{De Cat 
\& Aerts}{2002}]{DeCatAerts2002} De Cat P., Aerts C., 2002, A\&A, 393, 965 

\bibitem[\protect\citeauthoryear{Degroote et 
al.}{2010}]{Degroote2010} Degroote P., et al., 2010, Nature, 464, 259 

\bibitem[\protect\citeauthoryear{Dziembowski, Moskalik, \&
    Pamyatnykh}{1993}]{Dziembowski1993} Dziembowski W.~A., Moskalik P.,
  Pamyatnykh A.~A., 1993, MNRAS, 265, 588

\bibitem[\protect\citeauthoryear{Gautschy 
\& Saio}{1993}]{Gautschy1993} Gautschy A., Saio H., 1993, MNRAS, 262, 213 



\bibitem[\protect\citeauthoryear{Howell et al.}{2014}]{Howell2014} 
Howell S.~B., et al., 2014, PASP, 126, 398 

\bibitem[\protect\citeauthoryear{Kurtz et al.}{2015}]{Kurtz2015} 
Kurtz D.~W., Shibahashi H., Murphy S.~J., Bedding T.~R., Bowman D.~M., 
2015, MNRAS, in press (arXiv:1504.04245)

\bibitem[\protect\citeauthoryear{Lee, Neiner, 
\& Mathis}{2014}]{Lee2014} Lee U., Neiner C., Mathis S., 2014, MNRAS, 443, 1515 

\bibitem[\protect\citeauthoryear{Martins et 
al.}{2015}]{Martins2015} Martins F., et al., 2015, A\&A, 575, A34 

\bibitem[\protect\citeauthoryear{Mathis, Neiner, 
\& Tran Minh}{2014}]{Mathis2014} Mathis S., Neiner C., Tran Minh N., 2014, A\&A, 565, A47 

\bibitem[\protect\citeauthoryear{Moravveji et 
al.}{2015}]{Moravveji2015} Moravveji E., Aerts C., P\'apics P.~I., 
Triana S.\ A., Vandoren B., 2015, \aap, in press (arXiv:1505.06902) 



\bibitem[Neiner et al.(2012)]{Neiner2012} Neiner, C., Floquet, M., Samadi, R.,
et al.\ 2012, A\&A, 546, A47

\bibitem[\protect\citeauthoryear{Neiner et al.}{2011}]{Neiner2011} Neiner C.,
  Wade G., Meynet G., Peters G., 2011, IAUS, 272,

\bibitem[Pamyatnykh et al.(2004)]{Pamyatnykh2004} Pamyatnykh, A.~A., 
Handler, G., \& Dziembowski, W.~A.\ 2004, \mnras, 350, 1022 

\bibitem[P{\'a}pics et al.(2012)]{Papics2012} P{\'a}pics, P.~I., Briquet, M.,
Baglin, A., et al.\ 2012, A\&A, 542, A55

\bibitem[\protect\citeauthoryear{P{\'a}pics et al.}{2014}]{Papics2014}
  P{\'a}pics P.~I., Moravveji E., Aerts C., Tkachenko A., Triana S.~A., Bloemen
  S., Southworth J., 2014, A\&A, 570, A8

\bibitem[\protect\citeauthoryear{P{\'a}pics et al.}{2015}]{Papics2015}
  P{\'a}pics P.~I., Tkachenko A., Aerts C., Van Reeth T., De Smedt K., Hillen
  M., \O stensen R., Moravveji E., 2015, ApJ, 803, L25

\bibitem[\protect\citeauthoryear{Rogers et al.}{2013}]{Rogers2013} 
Rogers T.~M., Lin D.~N.~C., McElwaine J.~N., Lau H.~H.~B., 2013, ApJ, 772, 
21 

\bibitem[\protect\citeauthoryear{Samadi et al.}{2010}]{Samadi2010} Samadi R.,
  Belkacem K., Goupil M.~J., Dupret M.-A., Brun A.~S., Noels A., 2010, Ap\&SS,
  328, 253


\bibitem[Shiode et al.(2013)]{Shiode2013} Shiode, J.~H., Quataert,
E., Cantiello, M., \& Bildsten, L.\ 2013, MNRAS, 430, 1736

\bibitem[\protect\citeauthoryear{Sim{\'o}n-D{\'{\i}}az}{2015}]{SSD2015} Sim{\'o}n-D{\'{\i}}az S., 2015, IAUS, 307, 194 

\bibitem[\protect\citeauthoryear{Sim{\'o}n-D{\'{\i}}az 
\& Herrero}{2014}]{SSD2014} Sim{\'o}n-D{\'{\i}}az S. \& Herrero A., 2014, A\&A, 562, A135 


\bibitem[\protect\citeauthoryear{Sim{\'o}n-D{\'{\i}}az et 
al.}{2010}]{SSD2010} Sim{\'o}n-D{\'{\i}}az S., Herrero A., 
Uytterhoeven K., Castro N., Aerts C., Puls J., 2010, ApJ, 720, L174 

\bibitem[\protect\citeauthoryear{Tassoul}{1980}]{Tassoul1980} 
Tassoul M., 1980, ApJS, 43, 469 


\bibitem[\protect\citeauthoryear{Waelkens}{1991}]{Waelkens1991} Waelkens C.,
  1991, A\&A, 246, 453

\end{thebibliography}

\acknowledgments 
Part of the research included in this manuscript was based on
funding from the Research Council of KU\,Leuven, Belgium under grant
GOA/2013/012.  We thank an anonymous referee for constructive comments
which improved this manuscript.  

\end{document}